\documentclass[reprint,showpacs,preprintnumbers,amsmath,amssymb,prc,floatfix]{revtex4-1}
\usepackage{float}
\usepackage{color}
\usepackage{graphicx}
\usepackage{dcolumn}
\usepackage{bm}
\usepackage{xcolor}
\usepackage{comment}
\usepackage[colorlinks,citecolor=blue,linkcolor=red,anchorcolor=blue,filecolor=blue,urlcolor=blue]{hyperref}
\usepackage{multirow}
\usepackage{threeparttable} 
\usepackage{adjustbox} 
\usepackage{color}
\usepackage{longtable}
\usepackage{csquotes}{}

\begin{document}


\title{Reply to the \textquote {Comment on \lq Effect of density and nucleon-nucleon potential on the fusion cross section within the relativistic mean field formalism\rq }}

\author{M. Bhuyan$^{1,2}$}
\author{Raj Kumar$^3$}
\author{Shilpa Rana$^3$}
\author{D. Jain$^4$}
\author{S. K. Patra$^{5,6}$}
\author{B. V. Carlson$^7$}

\bigskip
\affiliation{$^1$Department of Physics, Faculty of Science, University of Malaya, Kuala
Lumpur 50603, Malaysia}
\affiliation{$^2$Institute of Research and Development, Duy Tan University, Da Nang 550000,
Vietnam}
\affiliation{$^3$School of Physics and Materials Science, Thapar Institute of Engineering
and Technology, Patiala 147004, India}
\affiliation{$^4$Department of Physics, Mata Gujri College, Fatehgarh, Punjab 140407, India}
\affiliation{$^5$Institute of Physics, Sachivalaya Marg, Sainik School, Bhubaneswar 751005,
India}
\affiliation{$^6$Homi Bhabha National Institute, Anushakti Nagar, Mumbai 400085, India}
\affiliation{$^7$Instituto Tecnol\'ogico de Aeron\'autica, S\~ao Jos\'e dos Campos 12.228-900,
S\~ao Paulo, Brazil}
\bigskip

\date{\today}
\bigskip 
\begin{abstract}
\noindent
In reply to the Comment made by M. V. Chushnyakova {\it et al.} on our paper [Phys. Rev. C \textbf{101}, 044603 (2020)], we argue that the calculations, results and conclusions of our paper remain valid. We have shown here the calculations for one reaction using the deformed densities and the R3Y nucleon-nucleon potential obtained within the relativistic mean-field (RMF) formalism. Suitable clarifications and justifications are given to address all the points raised in the Comment.
\end{abstract}

\maketitle

The main concern of the authors of comment \cite{chus22} on our paper \cite{bhuy20} is regarding the use of spherical densities for the considered target nuclei. First, we would like to clarify here that the spherical densities were used in \cite{bhuy20} as a simplifying assumption, as it was the first step towards implementing a new theoretical formalism, that is, the relativistic mean field (RMF) approach for the study of nuclear fusion. It is well-known that the deformations and orientations of both the fusing nuclei have significant effects on the interaction potential and, consequently, on the nuclear fusion. Moreover, in addition to nuclear shape degrees of freedom, nuclear fusion is also affected by numerous other factors such as nuclear shell corrections, pairing energy, incompressibility of nuclear matter, energy dependence of the interaction potential, mass, charge, and isospin asymmetry of the entrance channel \cite{mont17,toub17,canto20,das98,raj20}. The step-wise refinements are done in the theoretical models to study the role of various structure effects on nuclear fusion. Following the similar footing, in our recent studies \cite{bhuy22,rana22}, the in-medium effects were introduced in the relativistic R3Y nucleon-nucleon (NN) potential in terms of the density-dependent nucleon-meson couplings and the nuclear potential obtained using this density-dependent R3Y (DDR3Y) NN potential was further used to study the nuclear fusion. 

Furthermore, the authors of the Comment have also presented the calculations for $^{48}$Ca+$^{238}$U reaction using the 2pF densities along with Paris and Reid versions of the effective M3Y NN potential \cite{chus22}. It is worth noting here that we have used the RMF densities along with the relativistic R3Y effective NN potential in \cite{bhuy20} to obtain the nuclear potential, and the results were compared with the Reid version of the M3Y NN potential. Therefore, to study the effect of deformation within the same microscopic RMF formalism, here we present the calculations performed with the deformed RMF densities and the relativistic R3Y NN potential for $^{48}$Ca+$^{238}$U reaction at same center of mass energies as considered in \cite{chus22}. The quadrupole ($\beta_2=0.283$) deformation parameter for $^{238}$U target nucleus is obtained within the axially deformed RMF formalism for NL3$^*$ parameter set, and its effect is further included in the description of the nuclear densities through the radial vector. The capture cross-section obtained from $\ell$-summed Wong model \cite{bhuy20,bhuy22,rana22} for $^{48}$Ca+$^{238}$U reaction using the deformed RMF densities and R3Y effective NN potential is shown in Table \ref{tab}. The experimental data \cite{itkis04} and the calculated cross-section obtained by folding the Reid version of M3Y NN potential with deformed RMF densities are also given in Table \ref{tab} for the sake of comparison. We also note that the experimental data used in our paper was taken from the recent work of Wakhle {\it et al.} \cite{wak18}, and the original references therein. It can be clearly noticed from Table \ref{tab} that the cross section increases with inclusion of target quadrupole deformations, especially at the lower barrier energies. On comparing the calculated cross-sections with the experimental data, it is observed that for the case of spherical densities, the R3Y NN potential gives a much better match with the experimental data than the M3Y NN potential. The overlap between the experimental and theoretical results further improves on accounting for the quadrupole deformations of the target nuclei, and the R3Y NN potential still gives a better match than the Reid version of the M3Y potential. Here, we have taken the impact of only quadrupole deformations of the target nucleus into account. A more detailed study with the inclusion of more nuclear shape degrees of freedom in the nuclear potential obtained within the RMF formalism is under process and will be communicated shortly. Further, the comparison of results obtained for Reid M3Y NN potential with deformed RMF densities and those presented in Comment \cite{chus22} using the 2pF densities shows that there is a noticeable difference in the cross-section obtained using the two different approaches for the nuclear density distributions. All these observations infer that nuclear fusion depends upon various structural properties of both of the interacting nuclei, and systematic studies are required to explore the impact of one or another factor.  

\begin{table*}
\caption{Capture cross-section $\sigma$ (mb) for $^{48}$Ca+$^{238}$U reaction calculated using effective R3Y and M3Y NN potentials folded with deformed RMF densities. The experimental data from \cite{itkis04} is also given for comparison.}
    \centering
    \renewcommand{\arraystretch}{2}
    \begin{tabular}{llllll}
            \hline \hline
       $E_{cm}$ (MeV) & & 181 & 186 & 193 & 201  \\ \hline
        $\sigma_{exp}$ (mb) & & 0.0248 & 3.04 &	40.2 &	107 \\ 
        $\sigma_{sph}$ (mb) & R3Y-NL3$^*$ & 9.133 $\times 10^{-5}$ & 3.652 & 41.202 & 105.93 \\
                            & M3Y-Reid & 1.183 $\times 10^{-10}$ & 1.736 $\times 10^{-5}$ & 3.073 & 99.992 \\   
        $\sigma_{def}$ (mb) & R3Y-NL3$^*$ & 0.0294 &	3.698 &	42.844 &	105.93 \\
                            & M3Y-Reid & 0.0077 &	1.864 & 24.315 & 103.686\\  
     $ \frac{\sigma_{sph}} {\sigma_{exp}}$  & R3Y-NL3$^*$ & 0.004 &	1.201 &	1.025 &	0.990 \\ 
     & M3Y-Reid & 4.773 $\times 10^{-9}$	& 5.712$\times 10^{-6}$	& 0.076	& 0.935 \\ 
      $ \frac{\sigma_{def}} {\sigma_{exp}}$  & R3Y-NL3$^*$ & 1.187 & 1.217 &1.066 &	0.990 \\
      &  M3Y-Reid & 0.312 &	0.613 &	0.605 &	0.969 \\
      $ \frac{\sigma_{sph} (RMF)} {\sigma_{sph}(2pF)}$ & M3Y-Reid & 2.534 & 0.025 &	0.003 &	0.567 \\
      $ \frac{\sigma_{def} (RMF)} {\sigma_{def}(2pF)}$ & M3Y-Reid & 0.384 &	1.008 &	1.036 & 0.859 \\
       \hline       \hline
    \end{tabular}
    \label{tab}
    \end{table*}
 
Next, the author’s comment points out the use of the term \textquote{phenomenological} for M3Y NN-forces. In the original work, Ref. \cite{fold79}, the term “phenomenological” is used for the optical potential obtained using M3Y NN-potential. Both the M3Y and the R3Y attempt to be \textquote{microscopic} by folding two nuclear densities with what purports to be an effective nucleon-nucleon potential, which is \textquote{phenomenological}. One could argue (as we do) that the R3Y is more \textquote{microscopic} than the M3Y one, as it takes its parameters from a relativistic meson-exchange interaction that also describes the nuclear mean field rather than from a fit to non-relativistic G matrix elements.

The authors of the comment have also pointed out the concern regarding the references in our paper. The articles authored by Schiff {\it et al.} \cite{sch51,sch51a} are included as the bibliographical source in which a solution containing the extremely important non-linear $\sigma$-meson interaction terms is proposed. In Ref. \cite{sch51,sch51a}, Schiff develops the mathematical procedure that can be used to obtain an analytical solution for the field equation containing non-linear terms. The same procedure is also adopted to obtain the R3Y NN potential containing non-linear self-coupling terms \cite{sahu14}. Concerning the omission of other references on the M3Y potential, in most of the articles suggested by the authors of the comment, the density-dependent versions of M3Y NN potentials are adopted, which we have not used in our study \cite{bhuy20}. Although some of the references might have been missed by us, the references we considered relevant to our development were included. The same observation can be applied to the very important initial work by Walecka on the subject of relativistic mean fields. However, we point out that we did include in the references of our paper an important review of the subject by J.D. Walecka and B.D. Serot (see Ref. [53] of original article \cite{bhuy20}), published in Advances in Nuclear Physics in 1986.

In conclusion, these comments do not affect the correctness and relevance of our results and/or change the conclusions of our paper \cite{bhuy20}.

\section*{acknowledgments}
This work was supported by DAE-BRNS Sanction No. 58/14/12/2019-BRNS, FOSTECT Project Code: FOSTECT.2019B.04, FAPESP Project Nos. 2017/05660-0, FAPESP Projects No. 2017/05660-0, and by the CNPq - Brazil.



\end{document}